# The Role of Free Electrons in Ball Lightning Creation

## Herbert Boerner
Email: hf.boerner@gmx.de

## Abstract

After more than 180 years of research, ball lightning is still an unsolved problem in atmospheric physics. Since no progress can be expected without a controlled production of such objects in a laboratory, this report analyses a carefully selected subset of the observations, focusing on cases where the creation of ball lightning has been witnessed in order to identify the circumstances of their creation. Surprisingly, it was possible to establish that in many cases negative corona was involved as a precursor. Free electrons produced in a negative corona appear to be required for several processes, especially the creation of the visible plasma of the ball lightning and in forming a receiving antenna for the electromagnet pulse of the return stroke of the initiating cloud-ground lightning. In a different line of arguments, localized electromagnetic structures, being special solutions of Maxwell's equations, were identified as the most likely model for the physical nature of ball lightning. The antenna required to produce such a structure can also be due to the free electrons. The free electron hypothesis allows outlining further actions in terms of data collection, computer simulation and experiments.

## Introduction

Ball lighting has been observed since antiquity. Compilations of well described observations contain accounts which go back to the middle of the 17[th] century [1], and new observations are routinely recorded [2].

Yet after more than 180 years of research [3] ball lightning is still an unsolved problem in atmospheric physics [4]–[6]. There is no consensus on the physical nature of these objects, and so far, there are no experiments which have created objects that match the observed characteristics of ball lightning.

Recently, significant progress has been achieved from correlation of ball lightning observations with data from lightning detection networks [6], [7], establishing a cause – effect relation between these two phenomena.

Since no progress can be expected without the controlled production of such objects in the laboratory, it is of importance to define the environmental circumstances that lead to the creation of ball lightning.

The few cases where the start of ball lightning objects was witnessed are the main sources of information, where one can learn something about the physics of the ball lightning creation process.

The aim of this report is the collection of all information that might be relevant, even if only remotely, for the creation of such objects in the laboratory and to outline possible experiments.

## Observations

The creation of ball lightning has been observed in different situations:

- Ball lightning objects can branch off the channel of linear lightning. This is the only case, where photographic evidence is available [6] (case 2).
- The ball lightning object can be produced from a conductor, sometimes a metallic fence, which had been hit by lightning [6] (case 23).
- It can appear "out of thin air", far away from lightning channels and conductors. This behavior is well documented [1], [6] (case 1).
- Near or in aircraft during flight.

Since Brand's analysis of observations [1] it has been established that the creation of ball lightning is usually associated with linear lightning, or at least with thunderstorm conditions. An exception are the observations in aircraft, where the correlation with external causes is less clear.

For the further discussion, it is assumed, that there is only one type of ball lightning and that the observations do not describe several similar phenomena. Considering the current state of understanding of this phenomenon, this appears to be a reasonable starting point, observing the principle of simplicity. This assumption is also supported by the analysis of a European database [8], which indicates the existence of a core phenomenon.

Since the reports by casual observers are the only source for information on this subject, especial care is required to establish an information base that is as dependable as possible. Therefore, only the following types of information have been used in this analysis:

- Characteristics observed repeatedly and independently over a long period of time, which are therefore likely to be correct.
- Single events which are very well documented and provide detailed information.
- Accounts by people with professional training.

Not all situations where ball lightning is created provide useful information on the production process and for the design of experiments. The creation of a lightning channel is one example, and another one is the appearance of ball lightning in aircraft. The lightning channel is of course an ample source of energy, but the actual situation can neither be analyzed properly, nor can it be recreated faithfully in a laboratory. The situation in an aircraft poses different problems: how such an object can be created in the Faraday cage of a modern aircraft is hard to understand.

The situation which imposes the most stringent constraints on the physical causes is the creation of ball lightning in air at a distance from the lightning channel.

A lightning stroke can act at distance only via three mechanisms:

- By the quasi-static electric field, either of the approaching leader or the charge in the thundercloud,
- by the electromagnetic pulse (EMP) of the return stroke,
- or by the induction due to the current change of the return stroke.

Induction can be excluded in the cases where the distance was of the order of kilometers.

An example of production of ball lightning at a large distance from the lightning is the Neuruppin case, where at least 11 ball lightning objects were created by a positive CG flash with an exceptional strength of 370 kA peak current [6], [9]. The lightning was located by the BLIDS system more than five kilometers east of the region, where the ball lightning objects were observed. Especially interesting are two observations, where the appearance of ball lightning objects was seen inside houses. The two objects appeared in a living room and in a work shed. Both objects were very bright, but short-lived. Similar observations have been reported several times:

- A ball lightning object appeared during a thunderstorm inside a room with a large window [10].
- Report by Turner [6] (case 12). One object appeared indoors under a Plexiglas skylight above a round brass table with a diameter roughly matching the diameter of the table. The object was very long-lived. The observer reported a high electric field.
- Various cases are mentioned in Brand's book: cases number 3 and 4 in a room, number 89 over the plate of the stove, 99 from the stove, 117 above a table, 137 along the lamp, 189 in a petroleum lamp.

The electric field of a thundercloud can produce corona discharges at elevated objects like masts, antennas, or spires, which can become visible in low-light conditions if the discharge is strong enough. Such a visible corona is called St. Elmo's fire. In Neuruppin, visible corona was observed on a metallic sieve resting on a wheelbarrow [9] (witness 7).

Leaders approaching ground also create a rising electric field, which first leads to corona and then to the creation of streamers, developing into an upward connecting leader that finally establishes the contact to ground.

The resulting return stroke, which discharges the leader's charge, produces the electromagnetic pulse (EMP).

For ball lightning objects created far away from a lightning channel, only the corona and the EMP can be the factors initiating and driving the production process of ball lightning. It is therefore essential to take a closer look at the properties of the corona discharge and its interaction with the EMP.

# Positive lightning and ball lightning

In Neuruppin, the multiple ball lightning objects were created by a very strong positive lightning, whose maximum current of 370 kA was at the upper range of observed lightning currents. Obviously, the conditions created by this super bolt were very favorable to produce ball lightning.

There is now considerable evidence that positive cloud-ground lightning has a much higher probability of creating ball lightning than negative CG lightning [6], [7], [11]. The analysis of Keul and Diendorfer points at roughly a factor of 10 between the probability of positive CG versus negative CG lightning, although negative lightning of rather moderate strength clearly also generates ball lightning objects. The main difference between positive and negative CG lightning is that they create different types of coronas at ground: positive lightning will produce negative corona, whereas negative lightning will produce positive corona. It appears likely, that the different properties of negative and positive corona are the reason behind the difference in ball lightning production probability.

# Negative corona

Corona discharges develop at sharp points, where the electric field is enhanced, and an ionization avalanche can develop. If the point is positively charged, i.e., it is the anode, the electrons created in the ionization region ahead of the point are collected by the anode and the positive ions are moving away in the electric field. If the point is negatively charged, it is acting as the cathode and the free electrons created in the ionization region are moving away from the point. Since oxygen molecules have an electron affinity of about 0.45 eV (Table 1), free electrons become quickly attached within less than 100 nsec [12].

Therefore, their velocity is reduced to the drift velocity of the negative ion, which is about three orders of magnitude lower.

Since the negative ions move slowly, a considerable space charge is built up and, in the case of a negatively charged tip, it reduces the electric field stopping avalanche creation. When this negative space charge diffuses away, the field again becomes strong enough to create a new avalanche, and the whole process repeats itself. This leads to very regular pulses in the case of a negative tip, called Trichel pulses. Two competing processes are involved: ionization by electrons and electron capture, and the balance between them determines if the avalanche grows or not. The important factor is the strength of the electric field. If it is strong enough in the regions further away from the emissive tip, the discharge can propagate and form a tiny filament of conductive plasma which grows into the region of lower electric field, a streamer.

Only negative corona creates Trichel pulses. These pulses produce quite regular bunches of negative charge, which drift away under the influence of the electric field. In the case of the field of a thunderstorm cloud or an approaching leader, this gives the negative space charge a periodic structure in the vertical direction.

Streamers are conductive filaments with field enhancement at their tips, which enables them to grow into regions of lower electric field [13]. Negative and positive streamers are quite different [14]. The crucial difference with respect to ball lightning generation is that negative streamers require a more than two times higher field (about 9 kV/cm versus about 4 kV/cm) than positive ones for propagation. This means that positive CG lightning produces less and smaller streamers starting from the ground, a factor which also adversely influences the capability of lightning rods to attract positive leaders [6]. It is likely that normally streamers are created by an approaching positive leader, which then compete in terms of the available energy with the ball lightning formation. Only in rare cases, where no streamers could form, ball lightning objects are initiated. Case number 184 in Brand's book reports on such an event, where in a village in France a strong lightning created large streamers everywhere, except above a body of water where a ball lightning object was created. The water provided a flat, conducting surface where corona could develop, but which was less suitable for the initiation of streamers.

The negative oxygen ions will further react with other oxygen and nitrogen molecules, forming Ozone and nitrous oxides.

At lower field strengths, free electrons are almost absent in the negative space charge, but they can be created again if the electric field is high enough. Figure 1 shows the fraction of the negative charge that is due to free electrons. At 400 kV/m half of the negative charge is electrons, which were detached from the negative ions. The detachment is much easier than the ionization of neutral molecules, since the energies required are much lower (Table 1, Table 2).

**Table 1 - Electron affinity and mobility of negative ions (https://webbook.nist.gov/chemistry/)**

| Anion | Electron affinity [eV] | | Mobility [cm$^2$/Vsec] |
|---|---|---|---|
| | | | |
| $O^-$ | 1.462 | | |
| $O_2^-$ | 0.448 | | |
| $O_3^-$ | 1.899 | | |
| $N_2O_2^-$ | 3.351 | | 2.52 |
| $NO_3^-$ | 3.937 | | 2.14 |

Table 2 - Excitation and ionization energy of gases (https://webbook.nist.gov/chemistry/)

| Gas | First excitation energy [eV] | First ionization energy [eV] |
|---|---|---|
| H | 10.2 | 13.6 |
| $H_2$ | 10.8 | 15.9 |
| $N_2$ | 6.3 | 15.6 |
| $O_2$ | 7.9 | 12.1 |
|  |  |  |
| $H_2O$ | 7.6 | 12.7 |
| $CO_2$ | 10.0 | 14.4 |
| $SF_6$ | 6.8 | 15.6 |
| He | 19 | 24 |

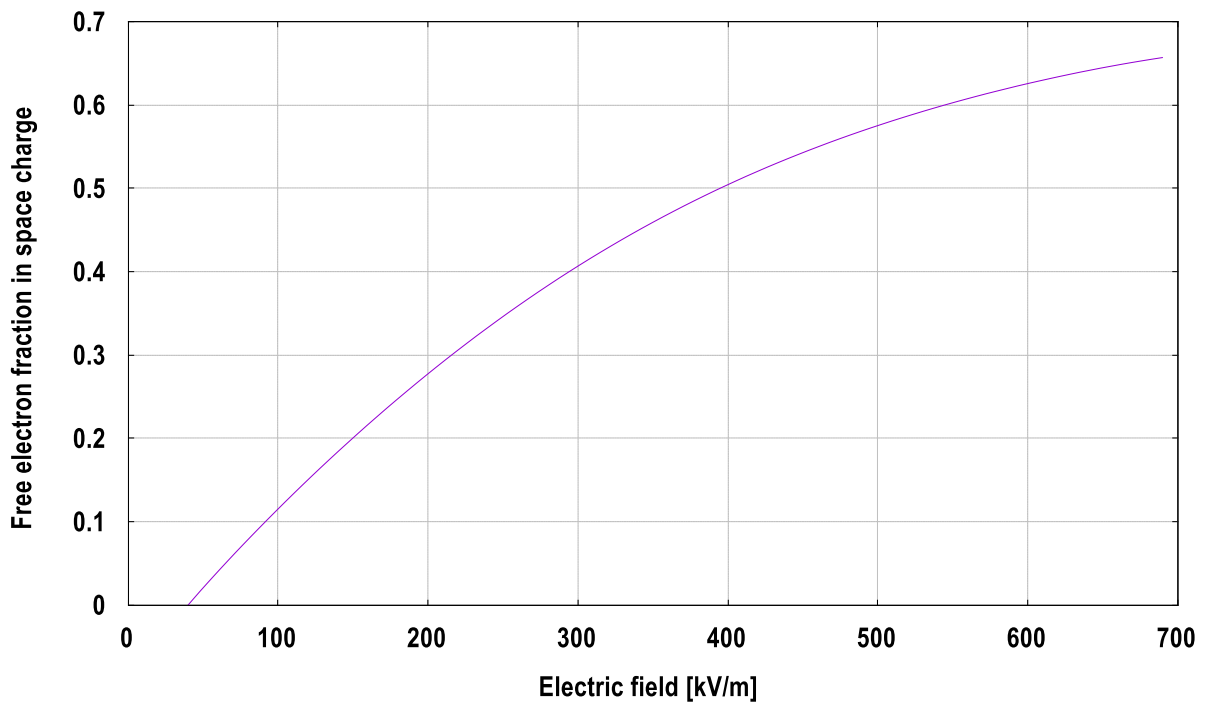

Figure 1 Fraction of free electrons in space charge, data from [12].

The sequence of events for a positive CG stroke is:

- The positive leader is approaching ground from the cloud above.
- When the electric field at ground is sufficient, a negative corona with Trichel pulses is produced.
- The negative oxygen ions form a periodic space charge, drifting vertically in the leader's field.
- When the electric field becomes larger, electrons are detached from negative oxygen, now creating a periodic structure of free electrons accelerating under the electric field of the leader.
- When some of the electrons have gained enough energy, they are able to ionize neutral gas molecules. The positive ions are left behind, and the multiplied electron bunch moves on.
- When no streamer discharge could be started, the process stops when the leader from above is connected to the ground, and the return stroke discharges the lightning channel.

- The electromagnetic pulse of the return stroke accelerates the free electrons in the space charge created by the corona.
- When the EMP stops, the electrons are attracted back to the positive ions created in their wake during acceleration upwards.
- The accelerated electrons emit electromagnetic radiation.

This scenario is textbook knowledge, containing no speculation. The only open question is how this sequence of events sometimes leads in the end to the creation of a ball lightning object.

The two sources of energy to produce ball lightning in these situations are the electric field and the EMP. Since the electric field is a weak source of energy, providing only about 40 J/m$^3$ at breakdown strength, the EMP is most likely the major source of energy for the ball lightning object. It is known, that the EMP of strong, especially positive lightning, interacts with the free electrons in the Ionosphere creating the ring-shaped luminous events called ELVES.

The free electrons of the negative corona are much closer to the source of the EMP than the electrons in the ionosphere and will therefore gain much more energy from it.

# Physical nature of ball lightning: the theory that fits the observations best

To get an idea how the structured electron cloud generated by the negative corona could possibly generate a ball lightning object, one needs to know what type of physical structure ball lightning probably is.

Since there is no agreement on the theory, one must pick the most likely one, which is the one that agrees best with the observed properties. Stenhoff remarks that until 1999 there was only one attempt to correlate observations with the theoretical models[1], but in [6] (chapter 11) the conclusion was reached that the theory that fits best is based on special solutions of Maxwell's theory [15]. This conclusion was mainly reached by the observation that ball lightning can pass through dielectric objects, like glass windows. Often the windowpanes are undamaged, but sometimes holes are punched into the panes by "cutting" out a piece of glass, which can be more or less round.

The passage through windows is well documented. The earliest one is case 192 by Brand in 1914, and the most recent one was in 2017 in Devon [6] (case 15), where a negative CG lightning hit the building adjacent to the observer's. Rakov [6] (case 17) lists one account, and [16] provides 43 accounts from Russia. Another report is from the Cavendish laboratory in Cambridge [6] (case 14).

This capability of ball lightning objects demonstrates that these objects cannot be entirely made of matter, they must primarily be composed of electromagnetic radiation, which produces the visible envelope of plasma. These exact solutions of Maxwell's equations are not electromagnetic waves propagating though space, but they have looped electric field lines of finite extent and a localized appearance in all three spatial dimensions [15]. These objects behave in fact more like particles than waves. The energy is stored in the electromagnetic field, and the stability is also provided by the field configuration, so no external reflector is needed. The configuration of the fields is not fixed, there are a huge number of possible structures [15], also including ones that are tangled and look like a "ball of yarn".

The visible plasma envelope is generated by the high electric fields of the EM structure, which accelerate free electrons such that air molecules become ionized and excited. Thus, the energy stored in the EM structure is slowly depleted, until the structure cannot be maintained anymore and the ball lightning object either just disappears or explodes or implodes with a noise. When such a structure moves through air, new plasma is created ahead of the object, whereas behind it the plasma recombines.

---

[1] The actual report could not be obtained by the author: Hubert, P. (1996) Novelle enquête sur la foudre en boule—analyse et discussion de résultats, Rapport PH/SC/96001, Centre d'Etudes Nucleaire, Saclay.

The conclusion that ball lightning objects are not composed of matter but of radiation, is also supported by an observation of the German physicist Walther Gerlach, who estimated the speed of the object he saw to 1200 m/sec [17], which is a speed that rules out any material object.

Cameron calls these objects "electromagnetic disturbances" [15], and proposes antenna configurations to produce them in the laboratory [18].

In the case of ball lightning objects created in open air, there is no metal antenna available, but the structure of the space charge of free electrons can function as such. How exactly the accelerated electrons can produce such an electromagnetic structure is of course an open question, but it is evident that free electrons are a necessary part of the process.

Earlier papers discussing this type of model are [19] and [20].

## Producing a discharge in air without electrodes

Another argument is also leading to the requirement of free electrons in air. Ball lightning objects are luminous, and this shows that at least their outer, visible parts are composed of a thin plasma.

In air at normal pressure, away from any electrodes, it is difficult to start a plasma. The initiation of a plasma requires free electrons, which can be accelerated by electric fields to energies which allow the ionization of more air molecules, starting an electron avalanche. At sea level, free electrons are created by radioactive decay of substances like radon or by secondary particles of cosmic radiation like muons. The production rate is about 10 ion pairs per cubic centimeter and second [21], leading to about 1000 ions per cubic centimeter.

The ionization of nitrogen and oxygen requires more than 10 eV (Table 1 and Table 2), so the electrical field required for electrical breakdown under normal pressure is 3 MV/m. It is nevertheless possible to start a plasma at much lower field strengths, which is demonstrated by the creation of plasmoids in a microwave oven. The electric fields in a microwave oven are about 2-3 kV/m,[2] which is three orders of magnitude below the breakdown field.

In such a low electric field, breakdown can only be achieved if an additional source of electrons is available, which can be supplied by a burning candle, a burning match, by fine carbon fibers, by a sharp metallic object, or a combination of those[3]. Flames of burning hydrocarbons have long been known to produce ions. The main process of the chemo-ionization is [22]

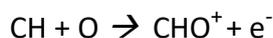

CH + O → CHO$^+$ + e$^-$

The visible flame is positively charged, as can be seen when, for example, a candle flame is burning in a strong electric field: the flame is drawn towards the negative electrode. Most of the free electrons from this reaction will be rapidly attached to oxygen, but not all. If the number of electrons is sufficiently high, there will be also enough electrons in the high-energy tail of the distribution, which can ionize air molecules, that ionization dominates recombination, and a plasma is started. Once initiated, the plasma is self-sustaining since the free electrons can absorb the microwaves and their full energy is coupled into the plasma. The hot plasmoid will rise to the top of the microwave oven and is therefore usually contained in a glass vessel by the experimenters.

---

[2] Estimated for a power of 1kW radiating into the oven cavity by the formula for Pointing's vector.
[3] For example: https://www.angelfire.com/electronic/cwillis/microwave.html

The increase in ionization due to burning hydrocarbons leads to a high concentration of free electrons, which then lowers the threshold for breakdown of air by three orders of magnitudes.

A similar effect is demonstrated by an interesting observation of Brand, whose significance has been overlooked by scientists until now. In his summary of ball lightning properties [1], [23], he states:

"They [ball lightning objects] are 'attracted' by the air in closed spaces; these balls enter the latter via an open window or door, and even through narrow cracks; however, they display a marked preference for entering via the flue gases of chimneys which have a better electrical conductance without self-induction, so that very often they emerge from the hearth and thus gain entrance into the kitchen."

Since chimneys are non-transparent, entering by this way cannot have been observed. Probably, the ball lightning objects were really created with high probability in places where hydrocarbons are burning, which is above the hearth or in an open fireplace.

A possible mechanism which may explain this surprising observation is the amplification of electron bunches. If free electrons move in a gas which contains many negative ions, the electrons can be stripped easily from the molecules, since the binding energies are much lower than the ionization energies. The flue gases work as an amplifying medium, which increases the number of electrons in the bunches of the Trichel pulses.

# Trichel pulses

To create an electromagnetic structure of the type described above, a specially tailored antenna is needed [18]. In open air, where no metallic antenna exists, a spatial distribution of the free electrons must function as an antenna. An unstructured blob of electrons will certainly be inadequate.

As described above, negative corona will run in a mode where Trichel pulses are created. In a homogeneous electric field, they will produce a periodic pattern of bunches of drifting negative ions. The mobility of negative ions in air at normal pressure is between 200 and 250 $mm^2$/Vsec [12]. The frequency of the Trichel pulses depends on the current the emissive point of the cathode can supply. For small corona currents, it is in the range of 100 Hz to about 10 kHz. The maximum frequency that could be obtained for Trichel pulses is about 3 MHz [24]. In Figure 2, the spacing of these ion bunches is shown, under the assumption of a constant electric field and for different frequencies. Typically, the spacing is in the submillimeter range, between 0.1 mm and 1 mm. The ion bunches thus form a fine spatial grating. If the electric field jumps to a value where free electrons are created, they will start from the ion bunches, retaining the periodic structure to a certain degree. After the EMP, which will deposit energy in the free electrons, the electron bunches will be attracted by the freshly created positive ions and the electromagnetic radiation from the accelerated, individual electron bunches will be interfering. It is therefore possible, that his unusual periodic arrangement of negative ions is the source of an electromagnetic radiation with wavelengths in the range of 0.1 to 1 mm, with corresponding frequencies of 300 GHz and 3000 GHz. The hypothesis, that Terahertz radiation is the source of ball lightning objects is certainly quite speculative, but it is supported by some observational facts. Ball lightning sometimes moves into keyholes, or it passes metallic mosquito screens [5] (20.2.4), which have a typical mesh size of one millimeter. The wavelengths of the electromagnetic structures must therefore be smaller than the openings in these metallic objects.

It should also be noted that devices like magnetrons, klystrons, or gyrotrons also work with bunches of electrons, but of course in a high vacuum. These devices deliver radiation continuously, whereas the radiation which may form the ball lightning object needs only to be produced in a very short burst, but then continues in a localized form.

In a realistic setting, not only one point will produce a corona, but there will be several emissive points, each running at a different current and frequency. In a case mentioned by Turner [6] (case 10) an unusually large ball lightning object occurred indoors above a metallic table, which was comparable in size to the ball. In this case, the whole surface of the table was acting as a cathode.

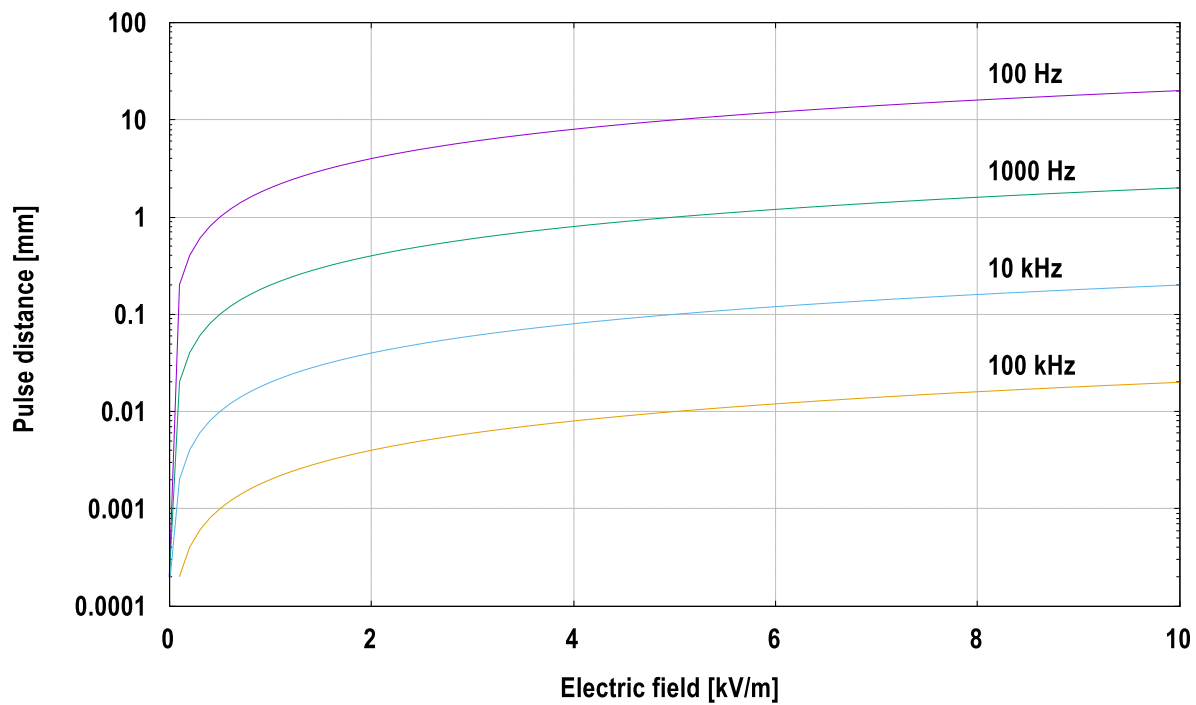

Figure 2  Spacing of Trichel pulses (for a mobility of 200 mm$^2$/Vsec)

# Ball lightning creation in situations not associated with positive CG lightning

The hypothesis that electrons from negative corona are the central mechanism for ball lightning creation explains the preference to positive CG lightning, but often ball lightning is observed to originate in other situations. In particular negative CG lightning of rather normal strength often produced ball lightning [7]. In such cases, the corona produced by the field at ground level of the stepped leader will be a positive one, but there are secondary effects which can lead to a negative corona. Lightning can hit metallic fences, open air telephone lines or power lines, which then will be charged negative. Negative corona can therefore be produced far away from the point where the lightning hits the conductor and ball lightning has often been observed to exit from such conductors, even from power sockets [6] (case 21, 23, 24).

Another possibility to create a negative corona by a negative CG stroke is by electrostatic induction. In the case of the unlucky Professor Richmann, who died during his attempts to duplicate Franklin's experiment, his equipment was an ungrounded lightning rod. The engraver named Sokolov described what happened during a thunderstorm: "Richmann was a foot away from the iron rod, when a pale ball of fire, the size of a fist, came out of the rod without any contact whatsoever. It went straight to the forehead of the professor, who in that instant fell back without uttering a sound." The upper end of the lightning rod will have sent out a positive upward connecting streamer, and therefore the lower end of the rod will have been negatively biased.

The occurrence of lightning channels is well documented, but it is obviously a rare phenomenon. Around the core of the lightning channel, which is highly conductive and usually negatively biased, a corona sheath develops, which stores the charge delivered by the lightning channel. In rare cases, ball lightning objects are created in this corona sheath. In these cases, the ball lightning objects often store a large amount of energy [6] (case 2).

Some curious observations of ball lightning become also understandable in the light of the negative charge hypothesis. In Italy, while filming a water fall, a ball lightning like object was accidentally recorded on video [25]. Waterfalls are copious sources of negative ions [26], so it is possible that they were the cause for this rare observation.

Ball lightning is also observed under conditions where the electric field is high, but where no thunderstorm is yet near, and no lightning is observed. Also, in some cases, ball lightning has been produced artificially and accidentally, such as by drawing an arc from a radio transmitter [6] (case 11, 18). In these situations, corona and consequently a space charge was obviously existing, but to what extent free electrons were present is not clear.

The observation of ball lightning inside modern, all-metallic airplanes presents the hardest problems. Sometimes, the ball lightning appears behind the cockpit window [27], or from the pilot's cabin, as in Jennings's report [6] (case 19). Obviously, the metallic body of the airplane acts as a Faraday cage, so no high electric fields can exist inside the plane, but the cockpit windows may be charged by the passage through charged regions in the clouds. Sometimes the plane was hit by lightning, as in the case reported by Jennings, but sometimes this is not the case [6], page 3).

# Summary

Free electrons appear to be fundamental for the creation of ball lightning objects. Because of their small mass, only electrons can easily be accelerated enough by electric fields so that they can start a plasma in air at normal pressure. They are also essential for absorbing the energy of the EMP of the return stroke, and only electrons can function as a transmitting antenna, which is needed to produce the localized electromagnetic structures. This hypothesis also allows understanding observations that hitherto were unexplained, for example the tendency of ball lightning objects to appear from fires or close to hearths. The negative ions in the flue gases provide an amplification medium for electron bunches originating from Trichel pulses of the negative corona. All situations where ball lightning was observed to originate are either associated with negative corona or at least with corona of unknown polarity.

The hypothesis that ball lightning is basically a localized electromagnetic structure based on a special solution of Maxwell's equations, is also compatible with the free electron hypothesis. Both hypotheses – free electrons as the fundamental means to create ball lightning and the localized electromagnetic structures – rest on independent lines of arguments and are well-supported by reliable observations.

The hypothesis that free electrons drive the creation of ball lightning objects offers a clear path for experimental verification. The situation, where ball lightning is created in air, can be recreated in the laboratory by producing a negative corona with Trichel pulses and then irradiating it with an electromagnetic pulse. Also, the situation where ball lightning is emanating from electric conductors, can be recreated in a laboratory. Nevertheless, it may be important to simulate the behavior of the negative space charge in Trichel pulses first in an appropriate model, to gain some insight into the settings of relevant parameters.

# Conclusion

The free electron hypothesis allows drawing several conclusions. For the collection of observational reports, it should become standard to check for circumstances that could provide negative space charge, like burning hydrocarbons, air ionizers, copying machines, laser printers and the like. So far, this type of information was only rarely reported. Also, the correlation with the type of the initial lightning is essential.

Computer simulations of negative corona, especially of Trichel pulses under pulsed excitation, offer the best chances to understand the processes prior to the self-organization which creates ball lightning objects. Such simulations will also be essential to define appropriate parameter settings for experiments.

For experiments, the most promising approach is probably one that recreates the situation that is produced by positive lightning. Alternatively, one could try to bias a conductor by a negative pulse of high voltage, imitating the cases where ball lightning emanates from conductors. Both approaches are well within the capabilities of normal physics laboratories.

Generally, both the process of self-organization that leads to the formation of ball lightning and the localized electromagnetic structures are of considerable scientific interest. The localized electromagnetic structures, which basically produce an electrodeless discharge in a gas, will probably have several interesting applications, which may also include nuclear fusion.